\documentclass[preprint,showpacs,preprintnumbers,
              superscriptaddress,floatfix]{revtex4}
\usepackage{amssymb}
\usepackage{graphicx}
\usepackage{dcolumn}
\usepackage{bm}
\usepackage{longtable}
\usepackage{multirow}% merge row
 \usepackage[colorlinks]{hyperref}

\newcommand{\dfrac}{\displaystyle\frac}

\begin{document}

\title{Tensor coupling effects on spin symmetry in anti-Lambda spectrum of hypernuclei}

\author{Chunyan Song}
\affiliation{State Key Laboratory of Nuclear Physics and Technology,
School of Physics, \\ Peking University, Beijing 100871, China}
\author{Jiangming Yao}
\affiliation{School of Physical Science and Technology, Southwest
University, Chongqing 400715, China} \affiliation{State Key
Laboratory of Nuclear Physics and Technology, School of Physics, \\
Peking University, Beijing 100871, China}
\author{Jie Meng}\email{mengj@pku.edu.cn}
\affiliation{School of Physics and Nuclear Energy Engineering,
Beihang University, Beijing 100191, China}
\affiliation{State Key
Laboratory of Nuclear Physics and Technology, School of Physics, \\
Peking University, Beijing 100871, China}
\affiliation{Department of
Physics, University of Stellenbosch, Stellenbosch, South Africa}

\date{\today}

%-------------------------------------------------------------------------------------------------------

\begin{abstract}
The effects of $\bar\Lambda\bar\Lambda\omega$-tensor coupling on the
spin symmetry of $\bar{\Lambda}$ spectra in $\bar{\Lambda}$-nucleus
systems have been studied with the relativistic mean-field theory.
Taking $^{12}$C+$\bar{\Lambda}$ as an example, it is found that the
tensor coupling enlarges the spin-orbit splittings of $\bar\Lambda$
by an order of magnitude although its effects on the wave functions
of $\bar{\Lambda}$ are negligible. Similar conclusions has been
observed in $\bar{\Lambda}$-nucleus of different mass regions,
including $^{16}$O+$\bar{\Lambda}$, $^{40}$Ca+$\bar{\Lambda}$ and
$^{208}$Pb+$\bar{\Lambda}$. It indicates that the spin symmetry in
anti-lambda-nucleus systems is still good irrespective of the tensor
coupling.

\end{abstract}

\maketitle

%-----------------------------------------------------------------------

 Spin symmetry and pseudo-spin symmetry in single particle
spectrum of atomic nuclei have been discussed extensively in the
literature. In atomic nuclei, there are a very large spin-orbit
splitting, i.e., pairs of single particle states with opposite spin
($j=\ell\pm \frac{1}{2}$) have very different energies. This fact
has allowed the understanding of magic numbers in nuclei and forms
the basis of nuclear shell structure\cite{MY.55}. More than thirty
years ago pseudo-spin quantum numbers have been introduced by
$\tilde\ell=\ell\pm 1$ and $\tilde{j}=j$ for $j=\ell\pm \frac{1}{2}$
and it has been observed that the splitting between pseudo-spin
doublets in nuclear single particle spectrum is an order of
magnitude smaller than the normal spin-orbit
splitting~\cite{AHS.69,HA.69}.

Pseudo-spin symmetry (PSS) is an important general feature in the
nuclear energy spectra and has been extensively discussed in the
framework of the relativistic mean-field (RMF)
theory~\cite{Bahri.92, Blokhin.95,Gin.97,MSY.98,MSY.99}. Since the
relation between the pseudospin symmetry and the RMF theory was
first noted in Ref.~\cite{Bahri.92}, the RMF theory has been
extensively used to study the pseudospin symmetry in the nucleon
spectrum. In Ref.~\cite{Blokhin.95}, it suggested that the origin of
pseudospin symmetry is related to the strength of the scalar and
vector potentials. Ginocchio took a step further to reveal that
pseudo-orbital angular momentum is nothing but the ``orbital angular
momentum" of the small component of the Dirac spinor, and showed
clearly that the origin of pseudo-spin symmetry in nuclei is given
by a relativistic symmetry in the Dirac Hamiltonian~\cite{Gin.97}.
The quality of pseudo-spin symmetry has been found to be related to
the competition between the centrifugal barrier and the pseudo-spin
orbital potential~\cite{MSY.98,MSY.99} within the RMF theory.

Recently, the possibility of producing a new nuclear system with one
or more anti-baryons inside normal nuclei has gained renewed
interest~\cite{BMS.02,Mishustin05,Friedman05,Larionov08,Larionov09}.
It motivates us to study the spin symmetry in the single
$\bar{\Lambda}$ spectrum, which can provide more information on the
antiparticles and their interaction with nuclei.

As the negative energy solutions to the Dirac equation are
interpreted as antiparticles under G parity transformation, the RMF
theory has been used to investigate the antinucleon spectrum, and a
well developed spin symmetry has been found in the antinucleon
spectrum~\cite{Zhou03}. The spin symmetry for anti-Lambda spectrum
in atomic nuclei has been examined and a better spin symmetry than
that in antinucleon has been reported in Ref.~\cite{Song09}.
However, the impurity effects of $\bar\Lambda$ and tensor coupling
effects were neglected there. Here in this work, the spin symmetry
in anti-Lambda hypernuclear system will be investigated with both
the impurity of $\bar\Lambda$ and tensor coupling included.

The tensor force has been discussed over many decades. Its
contribution to the spin-orbit splitting has been discussed by Arima
and Terasawa in terms of the second-order perturbation
~\cite{Arima60}. The importance of the tensor force for the nuclear
binding energy has been demonstrated in Ref.~\cite{Pudlineretal97}.
Recently, the tensor force was shown to have a distinct effect on
the evolution of the nuclear shell structure~\cite{Mao03,Otsuka05,
Colo07,Long07,Long08} and appropriate conservation of pseudo-spin
symmetry~\cite{Long061,Long07,Long10}. The importance of tensor
coupling effects in reducing the spin-orbit splitting of $\Lambda$
single particle energy spectrum has been extensively discussed in
the single-$\Lambda$ hypernuclei~\cite{Noble80,Jennings90,Yao08}.
Therefore, it is essential to examine further the spin symmetry of
$\bar{\Lambda}$ in $\bar{\Lambda}$-nucleus system with the presence
of $\bar\Lambda\bar\Lambda\omega$-tensor coupling with the RMF
theory.

For $\bar{\Lambda}$-nucleus system, the corresponding Lagrangian
density ${\cal L}$ can be written into two parts,
 \begin{eqnarray}
  \label{Eq:Lag}
   {\cal L} = {\cal L}_N + {\cal L}_{\bar{\Lambda}},
  \end{eqnarray}
where ${\cal L}_N$ is the standard Lagrangian density that has
already been extensively and successfully applied to ordinary
nuclei~\cite{Meng06}. The second part ${\cal L}_{\bar{\Lambda}}$ for
$\bar{\Lambda}$ hyperon with the
$\bar\Lambda\bar\Lambda\omega$-tensor coupling is given by,
 \begin{eqnarray}
  \label{Eq:Lag_Hyperon}
   {\cal L}_{\bar{\Lambda}}
    &=& \bar\psi_{\bar{\Lambda}}
      \left(  i\gamma^\mu\partial_\mu - m_{\bar{\Lambda}}
               - g_{\sigma \bar{\Lambda}} \sigma
               - g_{\omega \bar{\Lambda}} \gamma^\mu\omega_\mu
               \right)\psi_{\bar{\Lambda}}\nonumber\\
    &&- \frac{f_{\omega\bar{\Lambda}}}{4m_{\bar{\Lambda}}}
                 \bar\psi_{\bar{\Lambda}}
                 \sigma^{\mu\nu}\Omega_{\mu\nu}\psi_{\bar{\Lambda}}
 \end{eqnarray}
 where $m_{\bar{\Lambda}}$ is the mass of $\bar{\Lambda}$ and chosen as
$m_{\bar{\Lambda}}=1115.7$ MeV, $g_{\sigma \bar{\Lambda}}, g_{\omega
\bar{\Lambda}}$ are the coupling
 strengthes of $\bar{\Lambda}$ and $\sigma,\omega$ meson fields, and $f_{\omega \bar{\Lambda}}$ is
 the $\bar{\Lambda}\bar{\Lambda}\omega$ tensor coupling strength.
 The field tensor $\Omega_{\mu\nu}$ for the $\omega$-meson
 is defined as $\Omega_{\mu\nu}\equiv\partial_\mu\omega_\nu-\partial_\nu\omega_\mu$.

 With the mean field and {\it no-sea} approximations as well as the
 stationary condition for hypernuclear system, one obtains the Dirac equation
 for the $\bar{\Lambda}$,
 \begin{equation}
 \label{dirac}
 \left\{\mathbf{\alpha}\cdot \mathbf{p}+\beta
 (m_{\bar{\Lambda}}+S_{\bar{\Lambda}})
 +V_{\bar{\Lambda}}+T_{\bar{\Lambda}}\right\}\psi_{\bar{\Lambda}}
 =\epsilon_{\bar{\Lambda}} \psi_{\bar{\Lambda}},
\end{equation}
 where the scalar potential is $S_{\bar{\Lambda}}  =  g_{\sigma \bar{\Lambda}}
\sigma$, the vector potential $\displaystyle V_{\bar{\Lambda}}  =
g_{\omega \bar{\Lambda}} \omega_0$ and the tensor coupling potential
$\displaystyle
T_{\bar{\Lambda}}=-\frac{1}{2m_{\bar{\Lambda}}}\frac{f_{\omega
\bar{\Lambda}}}{g_{\omega \bar{\Lambda}}}i\bm\gamma\cdot\bm\nabla
V_{\bar{\Lambda}}$. The set of $\epsilon_{\bar{\Lambda}}$ values
forms the single-particle energy spectrum of $\bar\Lambda$.

According to the G-parity transformation, the coupling strengthes
for $\bar\Lambda$ and mesons are related to those for $\Lambda$.
Taking into account the many-body effects, an universal reduction
factor $\xi$ ($0<\xi\leq 1$) is introduced as that in
Refs.~\cite{Friedman05,Larionov08},
\begin{eqnarray}
g_{\sigma \bar{\Lambda}}&=&\xi g_{\sigma \Lambda},\\
g_{\omega \bar{\Lambda}}&=&-\xi g_{\omega \Lambda},
\end{eqnarray}
It has been found that the choice of $\xi=0.3$ is consistent with
the empirical $\bar{p}-A$ optical potential~\cite{Larionov08} and
will be used in the following studies. According to the
Okubo-Zweig-Iizuka (OZI) rule in naive quark model, the ratio
$\alpha(\equiv f_{\omega \bar{\Lambda}}/g_{\omega \bar{\Lambda}})$
of $\bar{\Lambda}\bar{\Lambda}\omega$ tensor coupling strength to
$\bar\Lambda$-$\omega$ coupling is taken as $\alpha=-1$
~\cite{Jennings90,Cohen91}.

With the restriction of spherical symmetry, the Dirac spinor of
$\bar{\Lambda}$ has the following form,
\begin{equation}
 \psi_{\bar{\Lambda}}(\bm{r}) = \displaystyle\frac{1}{r}
  \left(
   \begin{array}{c}
    i G_{n\kappa}(r)         Y_{jm}^{\ell}(\theta,\phi)        \\
    - F_{\tilde{n}\kappa}(r) Y_{jm}^{\tilde{\ell}}(\theta,\phi)
   \end{array}
  \right) ,
  \ \ j=l\pm \displaystyle\frac{1}{2},
 \label{eq:SRHspinor}
\end{equation}
where $Y_{jm}^{\ell}(\theta,\phi)$ are the spin spherical harmonics,
$G_{n\kappa }(r)/r$ and $F_{\tilde{n}\kappa }(r)/r$ form the radial
wave functions for the upper and lower components with $n$ and
$\tilde{n}$ radial nodes, and $\kappa = \langle 1 +
\mathbf{\bm{\sigma} \cdot \bm{\ell}} \rangle =
(-1)^{j+\ell+1/2}(j+1/2)$ characterizes the spin orbit operator and
the quantum numbers $\ell$ and $j$.

With the relation $\kappa(1+\kappa)=\ell(\ell+1)$, the Dirac
equation (\ref{dirac}) can be rewritten as a Schr\"{o}dinger-like
equation for the dominant component of Dirac spinor
$\psi_{\bar\Lambda}$,
\begin{eqnarray}
 \label{Gphi}
         -\displaystyle\frac{1}{2M_{+}^*}\left[\displaystyle\frac{d^{2}}{dr^{2}}+\displaystyle\frac{1}{2M_{+}^*}
\displaystyle\frac{dV_-}{dr}\displaystyle\frac{d}{dr}-\displaystyle\frac{\ell(\ell+1)}{r^2}\right]G_{n\kappa}(r)
\nonumber \\
 +\left\{\left(\displaystyle\frac{T(r)}{M_{+}^*}
-\displaystyle\frac{1}{4M_+^{*2}}\displaystyle\frac{dV_-}{dr}\right)\displaystyle\frac{\kappa}{r}
+\displaystyle\frac{1}{2M_{+}^*}[T^2(r)-T^\prime(r)]\right.\nonumber\\
\left.-
\displaystyle\frac{1}{4M_+^{*2}}\displaystyle\frac{dV_-}{dr}T+m+V_+\right\}G_{n\kappa}(r)
=\epsilon_{\bar\Lambda} G_{n\kappa}(r),
\end{eqnarray}
 with $V_{\pm}\equiv V_{\bar{\Lambda}}(r)\pm S_{\bar{\Lambda}}(r)$,
$2M_{\pm}^*\equiv m_{\bar{\Lambda}}\pm\varepsilon\mp V_{\mp}$,
$T(r)=-\displaystyle\frac{\alpha}{2m_{\bar{\Lambda}}}\displaystyle\partial_r
V_{\bar{\Lambda}}$, and $T^\prime(r)$ is the first derivative of
$T(r)$.

One notices in Eq.(\ref{Gphi}) that the total spin-orbit potential
(the term $\sim\kappa$, denoted as ``Total"), which determines the
energy difference between the spin-orbit partner states, is composed
of two terms. The first term is from the contribution of tensor
coupling ($\dfrac{T(r)}{M_{+}^*}$, denoted as ``Tensor"), and the
second term is the original spin-orbit potential from the derivative
of central potential
($-\dfrac{1}{4M_+^{*2}}\displaystyle\frac{dV_-}{dr}$, denoted as
``Central").

Recently, a new set of parameters for effective $\Lambda$-nucleon
interaction with $\Lambda\Lambda\omega$-tensor coupling, PK1-Y1
($g_{\sigma \Lambda}/g_{\sigma N}=0.580$, $g_{\omega
\Lambda}/g_{\omega N}=0.620$, $\alpha=-1$) is obtained by global
fitting the binding energies of single-$\Lambda$ hypernuclei in
different mass regions, based on the PK1 effective
interaction~\cite{Long04} for the nucleon part. The PK1-Y1 set is
shown great success in the description of both the single-$\Lambda$
binding energies and $\Lambda$ spin-orbit splittings~\cite{Lv09I}
and will be adopted in subsequent calculations.

\begin{figure}
 \includegraphics[width=6cm]{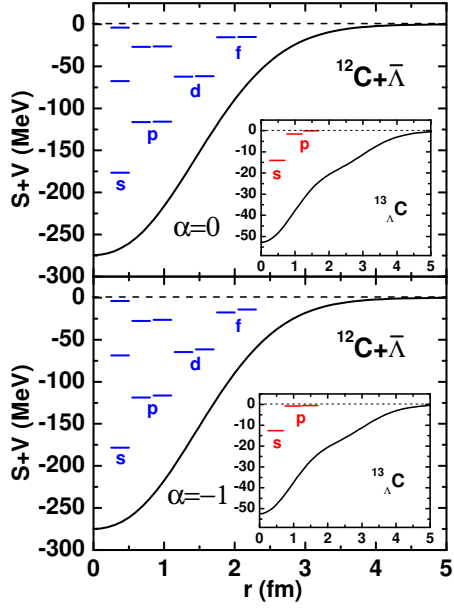}
\caption{ Single particle energies for $\bar{\Lambda}$
with($\alpha=-1$) and without($\alpha=0$) tensor coupling in
$^{12}$C+$\bar{\Lambda}$. For comparison, the insets show the
corresponding results for $\Lambda$ in $^{13}_{\Lambda}$C.}
 \label{fig1}
\end{figure}
Taking $^{12}$C+$\bar{\Lambda}$ as the first example, the effects of
tensor coupling on the spin symmetry in single-$\bar{\Lambda}$
energy spectrum are studied. Figure~\ref{fig1} shows the single
particle spectrum for $\bar{\Lambda}$ in $^{12}$C+$\bar{\Lambda}$.
In order to illustrate the tensor coupling effects on the spin-orbit
splittings, the single particle energy spectrum for $\bar{\Lambda}$
without($\alpha=0$) tensor coupling is also plotted in
Fig.~\ref{fig1}. For comparison, the corresponding results for
$\Lambda$ in $^{13}_{\Lambda}$C are given as well in the insets. It
shows clearly that the spin-orbit splittings of each spin doublets
for $\bar{\Lambda}$ are much smaller than those for $\Lambda$ if the
tensor coupling is not considered ($\alpha=0$). However, the
opposite phenomena occurs after taking into account the tensor
coupling ($\alpha=-1$), i.e., the spin-orbit splitting size becomes
very small for $\Lambda$ states as found in
Refs.~\cite{Jennings90,Cohen91,mares94,Ma96}, but significant for
$\bar{\Lambda}$ states.

   \begin{figure}[tbp]
 \includegraphics[width=8.5cm]{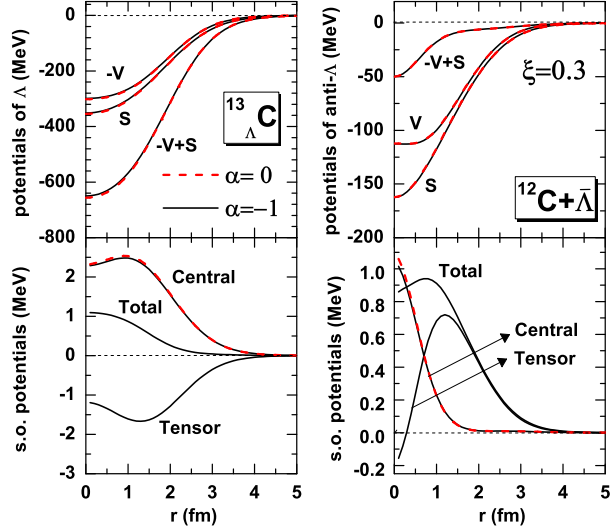}
\caption{The comparison of scalar, vector and total potentials
(upper panels) and spin-orbit potentials for $\Lambda$ in
$^{13}_{\Lambda}$C and $\bar\Lambda$ in $^{12}$C+$\bar{\Lambda}$
from the RMF calculations without ($\alpha=0$, dashed line)
 and with ($\alpha=-1$, solid line) the tensor coupling. }
 \label{fig2}
\end{figure}

We show in Fig.~\ref{fig2} the potentials of scalar $S(r)$, vector
$V(r)$ types and their difference $S(r)-V(r)$ (cf. Eq.(\ref{dirac}))
as well as the spin-orbit potentials (cf. Eq.(\ref{Gphi})) for
$\bar{\Lambda}$ in $^{12}$C+$\bar{\Lambda}$ from the RMF
calculations without and with the
$\bar\Lambda\bar\Lambda\omega$-tensor coupling. For comparison, the
corresponding results for $\Lambda$ in $^{13}_{\Lambda}$C are given
as well. As seen in Fig.~\ref{fig2}, the vector potential of
$\bar{\Lambda}$ changes its sign because of G-parity symmetry. The
derivative of the difference between the vector and scalar
potentials changes dramatically with the radial coordinate $r$ only
for $r$ smaller than 1.5 fm, which leads to the central part of
spin-orbit potential decreasing rapidly to zero at $\sim 1.5$ fm.
However, for $\Lambda$ in $^{13}_{\Lambda}$C, it is shown that the
difference between the vector and scalar potentials is quite large.
As the consequence, the corresponding central part of spin-orbit
potential is much larger than that for $\bar{\Lambda}$.

Of particular interesting is the onset of almost opposite phenomena
after taking into account the tensor coupling effects. The
contribution from tensor coupling (``Tensor") reduces the spin-orbit
potential for $\Lambda$, but enhances that for $\bar\Lambda$. These
effects can be observed on the splitting size of spin-orbit partner
states, as partly shown in Fig.~\ref{fig1}.

\begin{table}
 \tabcolsep=4pt
 \caption{Spin-orbit splittings of $\Lambda$ in $^{13}_{\Lambda}$C and
 of $\bar\Lambda$ in $^{12}$C+$\bar{\Lambda}$ from the RMF
 calculations without ($\alpha=0$) and with ($\alpha=-1$) tensor
 coupling. In the calculations with tensor
 coupling, the expectations of the spin-orbit potentials labled with ``Central", ``Tensor" and
 ``Total" in Fig.~\ref{fig2} are calculated with the dominant
 components in the Dirac spinors of spin doublets. Their differences
 are shown respectively in column ``$\Delta SOP$". All energies are in units of MeV.
 The experimental value of the spin-orbit splitting
 for $p_\Lambda$ states in $_{\Lambda}^{13}$C is $152\pm54\pm36$ keV~\cite{Ajimura01}.}
 \label{Tab1}
\begin{center}
\begin{tabular}{ccccccc}
\toprule
  & &\multirow{2}{*}{$\Delta E^{\alpha=0}$}& \multicolumn{3}{c}{$\Delta SOP$}&\multirow{2}{*}{$\Delta E^{\alpha=-1}$} \\ \cline{4-6}
  & &                     & Central& Tensor&Total&\\ \hline
  $_{\Lambda}^{13}$C        &$1p$ & 1.51   &  1.47             & -1.20   & 0.27   & 0.26      \\
 \hline
\multirow{4}{*}{$^{12}$C  +$\bar{\Lambda}$}&$1p$ &  0.64 & 0.64 & 1.85 & 2.49 &2.49 \\
                                           &$2p$ &  0.33 & 0.32 & 1.03 & 1.35 &1.37\\
                                           &$1d$ &  0.48 & 0.50 & 2.87 & 3.37 &3.37\\
                                           &$1f$ &  0.28 & 0.30 & 3.18 & 3.48 &3.47\\
 \hline\hline
\end{tabular}
\end{center}
\end{table}

 In Tab.~\ref{Tab1}, we give the values of spin-orbit splittings,
 \begin{equation}
 \Delta E=\epsilon(j=\ell-1/2) - \epsilon(j=\ell+1/2)
 \end{equation}
 for $\Lambda$ in $^{13}_{\Lambda}$C and  for $\bar\Lambda$ in $^{12}$C+$\bar{\Lambda}$
 from the RMF calculations both without ($\alpha=0$) and with ($\alpha=-1$) the tensor
 couplings. To show the tensor coupling effects on
 the splitting quantitatively, we calculate the expectations of the
 spin-orbit potentials with the dominant
 components in the Dirac spinors of spin doublets,
\begin{eqnarray}
SOP&\equiv&\int dr G(r)^2\left(\displaystyle\frac{T}{M_+^{*}}
-\displaystyle\frac{1}{4M_+^{*2}}\displaystyle\frac{dV_-}{dr}\right)\displaystyle\frac{\kappa}{r}\nonumber\\
&=&-\int dr
G(r)^2\displaystyle\frac{1}{4M_+^{*2}}\displaystyle\frac{d(V-S)}{dr}\displaystyle\frac{\kappa}{r}\nonumber\\
 &&-\int dr
G(r)^2\displaystyle\frac{1}{M_+^{*}}\displaystyle\frac{\alpha}{2m_Y}\partial_r
V\displaystyle\frac{\kappa}{r}.
\end{eqnarray}
where the first term is labled with ``Central", and the second term
is labled with ``Tensor", as indicated in Fig.~\ref{fig2}. The
difference of the expectations of total spin-orbit potentials
between the spin doublets ($\Delta SOP$) gives mainly the observed
spin-orbit splittings.

As seen in Tab.~\ref{Tab1}, the spin-orbit splitting for $p_\Lambda$
states of $^{13}_{\Lambda}$C is 0.26 MeV, which is in agreement with
the corresponding data $152\pm54\pm36$ keV~\cite{Ajimura01}. For
$\bar{\Lambda}$, the spin-orbit splittings of $1p, 2p, 1d, 1f$
states with the tensor coupling contribution are found to be
($1.37\sim3.47$) MeV, which is an order of magnitude larger than
those without the tensor coupling ($0.28\sim0.64$) MeV.

%It is noted that the spin-orbit splitting in the calculations
%without the tensor coupling, namely only given by the ``Central"
%part of spin-orbit potential, is almost the same as the ``Central"
%part of the spin-orbit splitting in the calculations with the tensor
%coupling.It indicates
%that the tensor coupling has negligible contribution to the
%``Central" part of spin-orbit potential through the rearrangement of
%mean-fields. However, the addition contribution from the tensor
%coupling to the spin-orbit potential of $\bar\Lambda$, corresponding
%to the ``Tensor" term, dominates the final spin-orbit splittings in
%the calculations with the tensor coupling. Table~\ref{Tab1} shows
%clearly that the ``Tensor" part of the spin-orbit splitting almost
%cancels the ``Central" part for $\Lambda$ states in
%$^{13}_\Lambda$C, but enhances that for $\bar\Lambda$ states greatly
%in $^{12}_\Lambda$C+$\bar\Lambda$.
It is noted that the spin-orbit splitting without the tensor
coupling is almost the same as the ``Central" part of the spin-orbit
splitting in the calculations with the tensor coupling. It indicates
that the tensor coupling has negligible contribution to the
``Central" part of spin-orbit potential through the rearrangement of
mean-fields. However, the addition contribution from the tensor
coupling to the spin-orbit potential of $\bar\Lambda$, corresponding
to the ``Tensor" term, dominates the final spin-orbit splittings in
the calculations with the tensor coupling. Table~\ref{Tab1} shows
clearly that the ``Tensor" part of the spin-orbit splitting almost
cancels the ``Central" part for $\Lambda$ states in
$^{13}_\Lambda$C, but enhances that for $\bar\Lambda$ states greatly
in $^{12}_\Lambda$C+$\bar\Lambda$.

\begin{figure}
 \includegraphics[width=8.5cm]{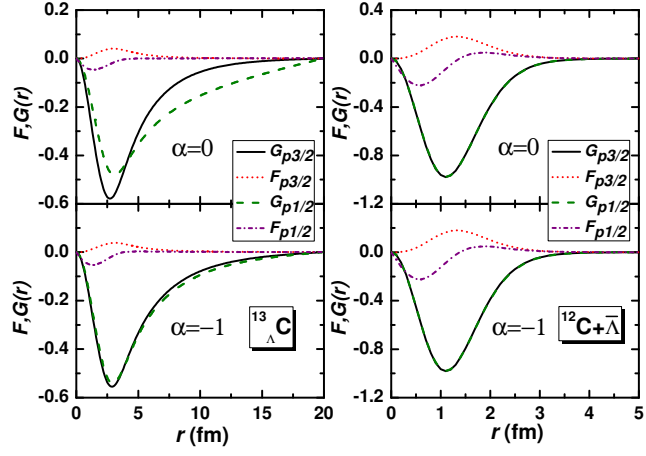}
\caption{ Radial wave functions for $p_\Lambda$ states in
$^{13}_{\Lambda}$C(left panel) and $p_{\bar\Lambda}$ states in
$^{12}$C+$\bar{\Lambda}$(right panel). In each case, the top panel
represents results without tensor coupling ($\alpha=0$) and the
lower part displays results with tensor coupling ($\alpha=-1$).}
 \label{fig3}
\end{figure}

In Fig.~\ref{fig3}, we plot the radial wave functions for
$p_\Lambda$ states in $^{13}_{\Lambda}$C and $p_{\bar\Lambda}$
states in $^{12}$C+$\bar{\Lambda}$ from the RMF calculations with
and without the tensor couplings. It shows clearly that the tensor
coupling effect is significant on the dominant components of Dirac
spinors for $\Lambda$ spin-orbit partner states. It recovers the
spin symmetry on the wavfunctions of $p_\Lambda$ spin-orbit partner
states. For $\bar\Lambda$, the same good spin symmetry is observed
from the calculations with and without the tensor couplings for the
dominant components of wavefunctions of spin-orbit partner states,
because the spin-orbit potential of $\bar\Lambda$ ($\sim1$ MeV) is
much smaller than the corresponding total potential
$V_{\bar\Lambda}+S_{\bar\Lambda}$ ($\sim280$ MeV). Therefore, the
changing of spin-orbit potentials due to the tensor coupling has
negligible influence on the final wavefunctions of $\bar\Lambda$
states.

\begin{table}
 \tabcolsep=4pt
 \caption{Spin-orbit splittings for different single-particle states of $\bar\Lambda$
 in $^{16}$O+$\bar{\Lambda}$, $^{40}$Ca+$\bar{\Lambda}$, and $^{208}$Pb+$\bar{\Lambda}$ from the
RMF calculations without ($\alpha=0$) and with ($\alpha=-1$) tensor
coupling.  All energies are in units of MeV.}
 \label{Tab2}
\begin{center}
\begin{tabular}
{ccccccc} \hline \hline
  & &\multirow{2}{*}{$\Delta E^{\alpha=0}$}& \multicolumn{3}{c}{$\Delta SOP$}&\multirow{2}{*}{$\Delta E^{\alpha=-1}$} \\ \cline{4-6}
  & &                     & Central& Tensor&Total&\\ \hline
                          &$1p$ & 0.39 & 0.40 & 1.48  & 1.88 & 1.88\\
                          &$2p$ & 0.23 & 0.22 & 0.89  & 1.11 & 1.12\\
$^{16}$O+$\bar{\Lambda}$  &$1d$ & 0.29 & 0.30 & 2.11  & 2.41 & 2.41\\
                          &$2d$ & 0.16 & 0.16 & 0.95  & 1.11 & 1.12\\
                          &$1f$ & 0.18 & 0.19 & 2.30  & 2.49 & 2.48\\
\hline
                          &$1p$ & 0.26 & 0.28 & 1.22  & 1.50 & 1.48\\
                          &$2p$ & 0.23 & 0.23 & 0.90  & 1.13 & 1.14\\
$^{40}$Ca+$\bar{\Lambda}$ &$1d$ & 0.08 & 0.09 & 0.73  & 0.82 & 0.82\\
                          &$2d$ & 0.17 & 0.18 & 1.29  & 1.47 & 1.46\\
                          &$1f$ & 0.05 & 0.05 & 0.70  & 0.75 & 0.75\\
\hline
                          &$1p$ & 0.12 & 0.15 & 0.64  & 0.79 & 0.76\\
                          &$2p$ & 0.15 & 0.15 & 0.60  & 0.75 & 0.76\\
$^{208}$Pb+$\bar{\Lambda}$&$1d$ & 0.00 & 0.00 & 0.05  & 0.05 & 0.05\\
                          &$2d$ & 0.03 & 0.03 & 0.27  & 0.30 & 0.30\\
                          &$1f$ & 0.00 & 0.01 & 0.06  & 0.07 & 0.06\\
 \hline \hline
\end{tabular}
\end{center}
\end{table}

The tensor coupling effects on the spin-orbit splittings for
$\bar{\Lambda}$ have been studied systematically for
$\bar{\Lambda}$-nucleus in different mass regions, including
$^{16}$O+$\bar{\Lambda}$, $^{40}$Ca+$\bar{\Lambda}$ and
$^{208}$Pb+$\bar{\Lambda}$ as shown in Tab.~\ref{Tab2}. The tensor
coupling effects on the spin-orbit splitting for
$^{16}$O+$\bar{\Lambda}$, $^{40}$Ca+$\bar{\Lambda}$ and
$^{208}$Pb+$\bar{\Lambda}$ are similar as those for
$^{12}$C+$\bar{\Lambda}$. Specifically, the spin-orbit splittings of
$\bar\Lambda$ in the calculations with the tensor coupling are found
to be ($1.12\sim2.48$) MeV in $^{16}$O+$\bar{\Lambda}$,
($0.75\sim1.48$) MeV in $^{40}$Ca+$\bar{\Lambda}$, and
($0.05\sim0.76$) MeV in $^{208}$Pb+$\bar{\Lambda}$, which are an
order of magnitude larger than those from the calculations without
the tensor coupling, i.e., ($0.16\sim0.39$) MeV in
$^{16}$O+$\bar{\Lambda}$, ($0.05\sim0.26$) MeV in
$^{40}$Ca+$\bar{\Lambda}$, and ($0\sim0.15$) MeV in
$^{208}$Pb+$\bar{\Lambda}$. Moreover, it is noted that the
spin-orbit splittings for $\bar{\Lambda}$ decrease with the mass
number $A$ no matter the tensor coupling is considered or not.

In summary, the tensor coupling effects on the spin symmetry of
$\bar{\Lambda}$ in several anti-Lambda-nucleus systems have been
studied in the RMF theory with the new effective hyperon-nucleon
interaction PK1-Y1 for $\Lambda$. The coupling strengthes of
$\bar\Lambda$ with meson fields are obtained using G-parity
transformation, where, as usual, an universal reduction factor
$\xi=0.3$, consistent with the empirical $\bar{p}-A$ optical
potential, is introduced to take into account the many-body effects.

For $^{12}$C+$\bar{\Lambda}$, the spin-orbit splittings with tensor
coupling are found to be ($1.37\sim3.47$ MeV for $\bar{\Lambda}$) an
order of magnitude larger than those without the tensor coupling
($0.28\sim0.64$MeV). The contribution from the tensor coupling has
the dominant contribution to the splittings. Since the mean-field
potentials for $\bar\Lambda$ are greatly large than the
corresponding spin-orbit potential, the dominant components of the
Dirac spinors for spin-orbit partner states are almost identical
irrespective of the tensor coupling. It indicates that the tensor
coupling effects on the wave functions are negligible for
$\bar{\Lambda}$. Therefore, the spin symmetry for $\bar{\Lambda}$ in
$^{12}$C+$\bar{\Lambda}$ system is still quite good even with the
consideration of the tensor coupling. Similar phenomena has also
been observed in $\bar{\Lambda}$-nucleus of different mass regions,
including $^{16}$O+$\bar{\Lambda}$, $^{40}$Ca+$\bar{\Lambda}$ and
$^{208}$Pb+$\bar{\Lambda}$, and the conclusion remains.

%-------------------------------------------------------------------------------------------------------
\section{Acknowledgements}

We would like to thank Avraham Gal for drawing our attention to
possible implications of the tensor coupling in Lambda and
antiLambda hypernuclei as well as illuminating discussions. This
work is partly supported by the National Key Basic Research
Programme of China under Grant No 2007CB815000, the National Natural
Science Foundation of China under Grant Nos. 10947013, 10975008 and
10775004, the Southwest University Initial Research Foundation Grant
to Doctor (No. SWU109011).

%-------------------------------------------------------------------------------------------------------

%-------------------------------------------------------------------------------------------------------


\begin{thebibliography}{99}

\bibitem{MY.55} M. G. Mayer and J. H. D. Jensen, Elementary Theory of Nuclear Shell
Struture (New York: Wiley) (1955).

\bibitem{AHS.69} A. Arima, M. Harvey, and K. Shimizu, Phys. Lett. B \textbf{30},
517 (1969).

\bibitem{HA.69} K. Hecht and A. Adler,
 Nucl. Phys. A \textbf{137}, 129 (1969).



\bibitem{Bahri.92} C. Bahri, J. P. Draayer, and S. A.
Moszkowski, Phys. Rev. Lett. \textbf{68}, 2133 (1992).

\bibitem{Blokhin.95} A. L. Blokhin, C. Bahri, and J. P. Draayer,
Phys. Rev. Lett. \textbf{74}, 4149 (1995).

\bibitem{Gin.97} J. N. Ginocchio, Phys. Rev. Lett. \textbf{78}, 436 (1997);
 Phys. Rep. \textbf{414}, 165 (2005); and references therein.

\bibitem{MSY.98} J. Meng, K. Sugawara-Tanabe, S. Yamaji, P. Ring, and A. Arima,
 Phys. Rev. C \textbf{58}, 628(R) (1998).

\bibitem{MSY.99} J. Meng, K. Sugawara-Tanabe, S. Yamaji, and A. Arima,
Phys. Rev. C \textbf{59}, 154 (1999).

\bibitem{BMS.02} T. B{\"{u}}rvenich, I. N. Mishustin, L. M. Satarov, J. A. Maruhn, H. St\"{o}cker, W. Greiner,
 Phys. Lett. B \textbf{542}, 261 (2002).

\bibitem{Mishustin05} I. N. Mishustin, L. M. Satarov, T. J. B{\"{u}}rvenich, H. St\"{o}cker, and W. Greiner,
 Phys. Rev. C \textbf{71}, 035201 (2005).

\bibitem{Friedman05} E. Friedman, A. Gal, and J. Mare$\check{s}$,
Nucl. Phys. A \textbf{761}, 283 (2005).

\bibitem{Larionov08} A. B. Larionov, I. N. Mishustin, L. M. Satarov, and W. Greiner,
Phys. Rev. C \textbf{78}, 014604 (2008).


\bibitem{Larionov09} A. B. Larionov, I. A. Pshenichnov, I. N. Mishustin, and W. Greiner,
 Phys. Rev. C \textbf{80}, 021601(R) (2009).


\bibitem{Zhou03} S. G. Zhou, J. Meng, and P. Ring,
 Phys. Rev. Lett. \textbf{91}, 262501 (2003).
\bibitem{Song09} C. Y. Song, J. M. Yao, J. Meng, Chin. Phys. Lett.
\textbf{26}, 122102 (2009).
\bibitem{Arima60} A. Arima and T. Terasawa, Prog. Theor. Phys. \textbf{23}, 115 (1960).

\bibitem{Pudlineretal97} B. S. Pudliner, V. R. Pandharipande, J. Carlson, S. C. Pieper, R.B. Wiringa,
Phys. Rev. C \textbf{56}, 1720 (1997).


\bibitem{Otsuka05}T. Otsuka, T. Suzuki, R. Fujimoto, H. Grawe and Y. Akaishi,
 Phys. Rev. Lett. \textbf{95}, 232502 (2005).

\bibitem{Colo07}G. Col\`{o}, H. Sagawa, S. Fracasso and P. F. Bortignon,
Phys. Lett. B \textbf{646}, 227 (2007)
\bibitem{Mao03} G. J. Mao, Phys. Rev. C \textbf{67}, 044318 (2003).
%\bibitem{Long062}W. H. Long, N. V. Giai, and J. Meng, Phys. Lett. B640, 150 (2006).

\bibitem{Long07} W. H. Long, H. Sagawa, N.V. Giai, and J. Meng, Phys. Rev. C \textbf{76}, 034314 (2007).
\bibitem{Long08} W. H. Long, H. Sagawa, J. Meng, and N. V. Giai, Europhys. Lett. \textbf{82}, 12001 (2008).
\bibitem{Long061}W. H. Long, H. Sagawa, J. Meng, and N. V. Giai, Phys. Lett. B \textbf{639}, 242 (2006).
\bibitem{Long10} W. H. Long, Peter Ring, J. Meng, N. V. Giai, and Carlos
A. Bertulani,  Phys. Rev. C \textbf{81}, 031302(R) (2010).






\bibitem{Noble80} J. V. Noble, Phys. Lett. B \textbf{89}, 325 (1980).

\bibitem{Dover84} C. B. Dover and A. Gal,
Progr. Part. Nucl. Phys., Vol. 12, p.171 (1984), ed. D. Wilkinson (
Pergamon Press, Oxford).

\bibitem{Jennings90} B. K. Jennings, Phys. Lett. B \textbf{246}, 325 (1990); M. Chiapparini, A. O. Gattone, and B. K.
Jennings, Nucl. Phys. A \textbf{529}, 589 (1991).


\bibitem{Yao08} J. M. Yao, H. F. L\"{u}, G. Hillhouse, and J. Meng, Chin. Phys. Lett. \textbf{25}, 1629
(2008).

 \bibitem{Meng06} J. Meng, H. Toki, S. G. Zhou, S.Q. Zhang, W. H. Long, and L. S. Geng,  Prog. Part. Nucl. Phys. \textbf{57}, 470
 (2006).
\bibitem{Cohen91} J. Cohen and H. J. Weber, Phys. Rev. C \textbf{44}, 1181 (1991).
\bibitem{Long04} W. H. Long, J. Meng, N. VanGiai, and S. G. Zhou,
Phys. Rev. C \textbf{69}, 034319 (2004).
\bibitem{Lv09I}H. F. L\"{u} \emph{et al.}, to be published.

%\bibitem{anp1} B .D. Serot \emph{et al},  Adv. Nucl. Phys. \textbf{16}, 1 (1986).

\bibitem{mares94}J. Mare\v{s}, B. K. Jennings, Phys. Rev. C \textbf{49}, 2472 (1994).

\bibitem{Ma96} Z. Y. Ma, J. Speth, S. Krewald, Baoqiu Chen, and A.
Reuber, Nucl. Phys. A \textbf{608}, 305 (1996).

\bibitem{Ajimura01} S. Ajimura \emph{et al.}, Phys. Rev. Lett. 86,
4255 (2001).
\end{thebibliography}
\end{document}